# Query Expansion Based on Crowd Knowledge for Code Search

Liming Nie, He Jiang*, Zhilei Ren, Zeyi Sun, Xiaochen Li

**Abstract**—As code search is a frequent developer activity in software development practices, improving the performance of code search is a critical task. In the text retrieval based search techniques employed in the code search, the term mismatch problem is a critical language issue for retrieval effectiveness. By reformulating the queries, query expansion provides effective ways to solve the term mismatch problem. In this paper, we propose Query Expansion based on Crowd Knowledge (QECK), a novel technique to improve the performance of code search algorithms. QECK identifies software-specific expansion words from the high quality pseudo relevance feedback question and answer pairs on Stack Overflow to automatically generate the expansion queries. Furthermore, we incorporate QECK in the classic Rocchio's model, and propose QECK based code search method $QECK_{Rocchio}$. We conduct three experiments to evaluate our QECK technique and investigate $QECK_{Rocchio}$ in a large-scale corpus containing real-world code snippets and a question and answer pair collection. The results show that QECK improves the performance of three code search algorithms by up to 64% in Precision, and 35% in NDCG. Meanwhile, compared with the state-of-the-art query expansion method, the improvement of $QECK_{Rocchio}$ is 22% in Precision, and 16% in NDCG.

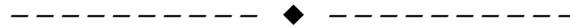

## 1 INTRODUCTION

CODE search is a frequent developer activity in software development practices, which has been a part of software development for decades [45]. As repositories containing billions lines of code become available [1], [3], [6], [41], the search mechanisms have evolved to provide better recommendation for given queries. On Google Code Search, a developer composes 12 search queries per weekday on average [39]. Meanwhile, developers search for sample codes more than anything else, 34% queries are conducted to find sample codes, and almost a third of searches are incrementally performed through query reformulation [39].

The performance of text retrieval based search techniques used in code search strongly depends on the text contained in queries and the code snippets (a method is viewed as a code snippet [21]). The term mismatch problem, also known as the vocabulary problem [13], is a critical language issue for retrieval effectiveness, as the queries given by users and the code snippets do often not use the same words [10]. Meanwhile, the length of queries is usually short. Sadowski et al. report that the average number of words per query is 1.85 for the queries proposed to Google search for code [39]. Obviously, it is not an easy task to formulate a good query, which depends greatly on the experience of the developer and his/her knowledge of the software system [35]. To solve the vocabulary problem, the query expansion methods provide some effective ways by reformulating the queries [10], [34].

In recent years, some query expansion based code search approaches are presented. For example, Wang et al. [56] incorporate users' opinions on the feedback code snippets returned by a code search engine to refine result lists. Hill et al. [38] suggest alternative query words by calculating the frequencies of co-occurring words with the words in the queries. Lu et al. [28] propose a query expansion method denoted as $P_{WordNet}$ by leveraging the Part-Of-Speech (POS) of each word in queries and WordNet [29] to expand queries. Lemos et al. [24] automatically expand test cases based on WordNet and a code-related thesaurus.

In this paper, we propose Query Expansion based on Crowd Knowledge (QECK) to improve the performance of code search. Specifically, QECK retrieves relevant Question & Answer (Q&A) pairs in a collection extracted from Stack Overflow as the Pseudo Relevance Feedback (PRF) documents for a given free-form query, identifies the software-specific words from these documents, and generates an expansion query by adding words to the original query. The advantages of QECK are three fold. First, it automatically generates expansion queries without human intervention, as QECK employs PRF to automatically generate expansion queries. Second, it generates high quality PRF Q&A pairs by considering textual similarity and the quality of both questions and answers. Third, it identifies software-specific words from Q&A pairs by *TF-IDF* weighting function.

The underlying idea behind QECK is utilizing the software-specific words contained in Q&A pairs to

• L. Nie, H. Jiang, Z. Ren, Z. Sun, and X. Li are with School of Software, Dalian University of Technology, Dalian, China. E-mail: limingnie@mail.dlut.edu.cn; jianghe@dlut.edu.cn; zren@dlut.edu.cn; sunzeyidlut@gmail.com; li1989@mail.dlut.edu.cn.

further improve the possibility of searching relevant code snippets. In Q&A pairs, the questions and answers, denoted as posts on Stack Overflow, are submitted and voted by developers. Therefore, the Q&A pairs contain useful knowledge about software development, which is called crowd knowledge in our study. The knowledge can be extracted in the form of software-specific words [52], [59]. Obviously, these software-specific words are more useful for software engineering tasks than the general words of WordNet used in previous studies [24], [28].

Pseudo relevance feedback is one of a local query expansion approaches, and the classic Rocchio's model is the implementation of pseudo relevance feedback in information retrieval. We incorporate QECK into the classic Rocchio's model, and propose QECK based code search method denoted as $QECK_{Rocchio}$. To evaluate the effectiveness of QECK and investigate the performance of $QECK_{Rocchio}$, we explore three Research Questions (RQs) in three experiments, respectively. These experiments are conducted on a Q&A pair collection containing 312,941 Q&A pairs labeled with the "android" tag, and a real-world code snippet corpus containing 921,713 code snippets extracted from 1,538 open source app projects on the Android platform. A code snippet refers to a method in Java files of app projects [21].

Three RQs and their conclusions are listed as follows.

*RQ1: Whether QECK can improve the performance of code search algorithms?*

We employ three code search algorithms to verify the effectiveness of QECK by comparing the recommendation performance before and after QECK is applied. From comparative results in the experiment, we verify that our QECK technique can indeed improve the retrieval performance for code search. Specifically, QECK improve the performance of three code search algorithms by up to 64% in Precision, and 35% in NDCG.

*RQ2: How the parameters affect the performance of QECK?*

For the parameters (i.e. the number of PRF documents and the number of expansion words) in QECK, we further study the influence of parameters variation on performance of QECK. As it is a time-consuming task to label relevant scores for code snippets, we only discuss the situation when we fix a parameter and explore the trend of performance for each code search algorithm by varying another parameter. Our results indicate that: after employing QECK, the performance of three algorithms are generally better, there is a unique optimal value of performance for each code search algorithm, and too many or too less expansion words is not desirable. Based on the results, we recommend that, in QECK, the default value for the number of PRF documents is 5, and the default value for the number of expansion words is 9.

*RQ3: Whether our code expansion based code search method, $QECK_{Rocchio}$, is better than the state-of-the-art method?*

We compare $QECK_{Rocchio}$ against $P_{WordNet}$, a state-of-the-art query expansion method [28]. The experimental results show that $QECK_{Rocchio}$ is a better method to aid mobile app development than the comparative method. Specifically, for Precision, the improvement is 22%, and for NDCG, the improvement is 16%.

This paper makes the following contributions:

- We propose QECK, a novel technique leveraging crowd knowledge on Stack Overflow to improve the performance of code search algorithms.
- We explore the performance and identify the effectiveness of QECK by three code search algorithms and a comparative method in terms of Precision and NDCG.
- We construct a Q&A pair collection from Stack Overflow and a code snippet corpus from open source app projects.

Next section outlines the background of our study. Section 3 elaborates our technique. Section 4 provides details about experimental setup. Experimental results and analysis are presented in Section 5. Section 6 states the threats to validity. The related works are shown in Section 7. In Section 8, we conclude this paper and introduce the future work.

## 2 BACKGROUND

In this section, we discuss the query expansion approaches, and elicit our query expansion based on crowd knowledge (QECK) technique.

The query expansion approaches, either fully automatically or with the help of users in the loop, contain two major classes: global approaches and local approaches [26], [58]. Here, we will mention the efforts on both of them, whereas we concentrate on the pseudo relevance feedback, a successful local approach used in our paper.

Global approaches mainly refer to query expansion/reformulation with a thesaurus, like WordNet [26]. The queries can be automatically expanded with related words and synonyms from the thesaurus. Specifically, WordNet is a general purpose lexical database to compute the semantic distance between two words. Sridhara et al. [43] show that the general English-based similarity measurements of WordNet could not effectively suggest similar words in software engineering context, as it does not contain many software-specific words, and the semantic meaning stored in WordNet is often different even though these words exist.

In the software engineering community, there are some efforts to automatically build a word similarity resource. For example, Yang and Tan [59] infer semantically related words by leveraging the context of words in software source code. Howard et al. [20]

seek to find similar verb pairs by leveraging the comments of methods, programmer conventions, and method signatures. Tian et al. [52], [53] build a software-specific WordNet like resource by leveraging the textual contents of posts on Stack Overflow. Although these resources can be employed to expand the words of queries, they omit the context of a query, which does not view a query as a whole [26].

Local approaches expand a query according to the documents initially appearing to match the original query, which mainly refer to relevance feedback and pseudo relevance feedback [26], [47]. Specifically, Relevance Feedback (RF) need to leverage user' marks on the RF documents as an initial set of results [56]. In contrast, to automate the manual part of relevance feedback, Pseudo Relevance Feedback (PRF), also known as blind relevance feedback, provides an approach for automatic local analysis [26]. Typically, this method assumes that a fixed number of top ranked documents are relevant to the original query, and extracts a set of potentially useful words from those documents and adds them to the query, which is then used to retrieve the final set of documents. This process is also called the classic Rocchio's model. The two following issues are mainly addressed in typical PRF approaches: retrieval of good quality feedback documents and identification of useful words [10].

In this paper, we propose query expansion based on crowd knowledge, one variation of PRF, to improve the performance of code search. QECK leverages the Q&A pairs to expand the queries. Different from the typical PRF, the initial set of results of QECK comes from a Q&A pair collection extracted from Stack Overflow rather than the code snippet corpus. Stack Overflow [7] is a popular question answering site, which provides a platform for developers to help others by asking and answering questions [52]. With about 4.8 million users and 11 million questions until November 2015, Stack Overflow is an enormous knowledge base. Most of the posts (questions or answers) submitted by users on Stack Overflow are related to software development. Leveraging the posts and their scores voted by crowd, QECK extracts software-specific expansion words to generate expansion queries.

## 3 OUR TECHNIQUE

This section shows the steps of $QECK_{Rocchio}$, and the construction of the Q&A pair collection and the code snippet corpus employed in our study. We also provide some details about the weighting of expansion words in the PRF documents.

### 3.1 Steps of $QECK_{Rocchio}$

In this study, for exploring the effectiveness of our QECK technique, we incorporate QECK into classic Rocchio's model to generate a code search method denoted as $QECK_{Rocchio}$. The Rocchio's model incorporates pseudo relevance feedback into the information retrieval process [41]. Fig. 1 shows the overall structure of $QECK_{Rocchio}$ containing three modules: *QA Pairs Search Engine, Expansion Words selector, and Code Snippets Search Engine*. The input of our method is an original query $q$, a Q&A pair collection, and a code snippet corpus. The output is a ranked list with top-k code snippets.

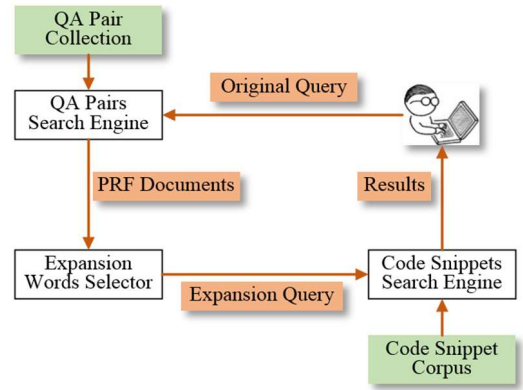

Fig. 1. Overall structure of QECK based Rocchio's model

The following three steps show the process of our method in Fig. 1.

- *First-pass retrieval:* For the original query, we rank all Q&A pairs using a particular information retrieval model (e.g. BM25 model) [46] by the module *QA Pairs Search Engine*. The top-m Q&A pairs are identified as the *PRF documents*, which is denoted as $D_f$. The *i-th* ranked document in $D_f$ is denoted as $d_i$, which will be treated as relevant to the original query.

- *Word selection:* We identify useful expansion words from PRF Q&A pairs ($D_f$) by the module *Expansion Words selector*. An expansion weight $w(t, d_i)$ is assigned to each word $t$ in the set of $D_f$. According to the weights of words, top-n words are selected and added into the original query to generate the *Expanded Query $q_e$*.

- *Second-pass retrieval:* Finally, we rank all code snippets in corpus for the expanded query $q_e$ using the module *Code Snippets Search Engine*. Finally, the top-k code snippets related to the expanded query are recommended to developers as the *results*.

In the above steps, we hold the expectation that the selected expansion words within the feedback documents can bring more relevant code snippets in the second-pass retrieval. Within this framework, three aspects are the important parts, which are the searching of PRF Q&A pairs, selecting of expansion words, and the construction of the Q&A pair collection and the code snippet corpus. We detail them in the next parts.

### 3.2 Q&A Pair Collection

#### 3.2.1 Q&A pairs

On Stack Overflow, there are many questions posted by users. The tags added to questions reveal the types

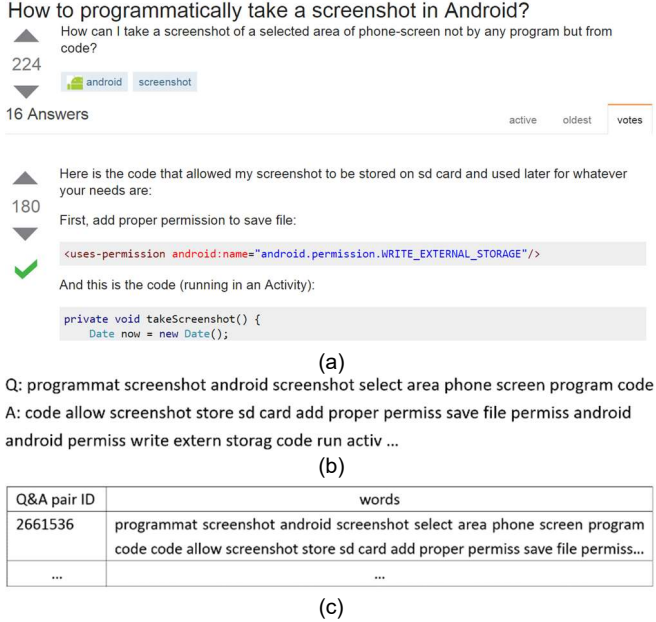

Q: programmat screenshot android screenshot select area phone screen program code
A: code allow screenshot store sd card add proper permiss save file permiss android android permiss write extern storag code run activ ...

(b)

| Q&A pair ID | words |
|---|---|
| 2661536 | programmat screenshot android screenshot select area phone screen program code code allow screenshot store sd card add proper permiss save file permiss... |
| ... | ... |

(c)

Fig. 2. An example of a Q&A pair (a), the words in Question (Q) and Answer (A) after text pre-processing (b), and the fields of a Q&A pair (c).

of these questions to help users find out what a question is about [31]. To generate Q&A pairs, we first select the questions with the "android" tags (a question on Stack Overflow can have up to five tags [46]), as the queries we employed in our experiments are about Android mobile app development. Next, we combine a question and one of their answers to generate a Q&A pair. However, for each of answers provided by users, it is not always fit to solve this question. To assure the quality of answers, only the answers labeled with "AcceptedAnswer" can be selected to form the Q&A pairs together with the corresponding questions. After the above steps, we achieve a Q&A pair collection related with Android development, which contains 312,941 Q&A pairs. In order to facilitate searching, we index this collection with Lucene [4].

Fig. 2 (a) shows an example of a Q&A pair for a programming task "*take a screenshot in Android*". In the part of question, this programming task contains a title, a detail description, two tags "*android*" and "*screenshot*". For this question, 224 votes and 16 answers are provided by crowd. Among the answers, only one of them is manually labeled as "accepted answer", which is marked with a tick. A part of this answer is shown in Fig. 2 (a). We combine the text in the question (the title and the description) and the accepted answer as a Q&A pair.

### 3.2.2 Indexing Q&A Pairs
We utilize Lucene [4], a popular implementation of BM25, to index and search Q&A pairs in the first-pass retrieval.

Before indexing the Q&A pairs, text pre-processing for Q&A pairs is required, which is an important process in the text retrieval [49]. In this process, first, the terms in the questions and answers are split by

Camel-case and separators (e.g., "_"). For example, the "MediaRecorder" can be split into "Media" and "Recorder". Second, we filter these words by removing the stop words. Finally, the remained words are handled by stemming [51]. After the above steps, a Q&A pair is now represented as a bag of words. Fig. 2(b) shows the pre-processed words in the question and the accepted answer.

In the process of indexing, each Q&A pair now represented by pre-processed words is stored as a document. Each document consists of a number of fields (i.e. the Q&A pair ID, words) as shown in Fig. 2(c). Now, we can search PRF Q&A pairs based on this index.

### 3.2.3 Searching Q&A Pairs
For searching the PRF Q&A pairs, following previous work [46], we consider the following two aspects. The first is the textual similarity, denoted by *Lucene score*, between Q&A pairs and queries calculated by BM25 similarity on Lucene. The other is the quality of Q&A pairs in terms of *SO score*. A *SO score* is a weighted mean value between the individual scores of its question and answer. The score of a post (question or answer) voted by crowd is regarded as a proxy for its quality. As their different natures of *Lucene score* and *SO scores*, we combine two scores by performing a normalization step to achieve the final score. Specifically, for the Q&A pairs returned in the first-pass retrieval, we normalize the *Lucene score* value of each pair and its *SO score* value in the range $[0,1]$ using min-max normalization technique [46]. For the i-th Q&A pair, the final score is calculated by following formulas:

$$ final\ score_i = \frac{L_i - min}{maxL - minL} + \frac{S_i - min}{maxS - minS} \quad (1) $$

$$ S_i = 0.7 * Sq_i + 0.3 * Sa_i \quad (2) $$

where, $L_i$ and $S_i$ are *Lucene score* and *SO score* of the i-th Q&A pair, respectively. The maximum and minimum of *Lucene score* and *SO score* among all Q&A pairs are represented with $maxL$ and $minL$, $maxS$ and $minS$, respectively. The symbols $Sq_i$ and $Sa_i$ refer to the values of question and answer in the i-th Q&A pair voted by crowd, respectively.

According to the final score, we rank the returned Q&A pairs in descending order and recommend top-m Q&A pairs as the PRF documents. Now, we can identify the useful expansion words from these PRF Q&A pairs [46].

### 3.3 Words Selection
A large number of approaches focus on finding good expansion words by weighting scores for words in the feedback documents [10]. Most of them are based on the assumption that the words that are most closely related to the query will have a comparatively higher probability of occurrence in the feedback documents. Following this general paradigm, various functions have been proposed to assign high scores to the words.

In our study, to weight the candidate words in a feedback document, we employ the traditional method, *TF-IDF* weighting function. *TF-IDF* is often used to determine the importance of a word for a particular document in the corpus [17]. We show the weighting of *TF-IDF* on Lucene as follows [4]:

$$w(t, d_i) = TF(t) * IDF(t) \qquad (3)$$

$$TF(t, d_i) = sqrt\big(tf(t, d_i)\big) \qquad (4)$$

$$IDF(t) = log\left(\frac{N}{df+}\right) + 1 \qquad (5)$$

where, term frequency, $tf(t, d_i)$, is the number of times the word $t$ appearing in a document $d_i$. The Inverse Document Frequency, $IDF(t)$, is the inverse of the number of documents in the corpus containing word $t$. $N$ is the total number of feedback documents. After weighting the words, top-n words can be identified as expansion words. Notably, as some words appearing in more than 25% of the documents in the collection are considered non-discriminating [15], [17], we eliminate these words in this process.

### 3.4 Code Snippet Corpus

We store the candidate code snippets in a corpus and index them on Lucene. The corpus contains 921,713 code snippets extracted from 1,538 open source app projects on the Android platform. The following three steps show the preparation for the code snippet corpus:

- Crawling the open source app projects

The code snippets in our experiments come from open source app projects on F-droid [2]. F-droid is a website with free and open source apps on the Android platform. Notably, among several versions of an app project, we select the latest version.

- Segmenting the Java files in these app projects

In order to collect candidate code snippets, following previous work [21], we utilize the tool Eclipse Abstract Syntax Tree (AST) to parse the Java files. Each Java file contains one or more methods. In the process of parsing, each method that containing the code text and comments is viewed as a code snippet.

- Indexing the code snippets on Lucene

Like the text pre-processing for Q&A pair, we also first process the code snippets before indexing them on Lucene. Then, each code snippet is stored as a document containing several fields (i.e. the name of code snippet and the words).

In the second-pass retrieval of *QECK_Rocchio*, we score each code snippet with the BM25 textual similarity between the expanded query and this code snippet. According to the scores, we select top-k code snippets as the final recommendation results.

## 4 EXPERIMENTAL SETUP

In this section, based on three research questions (RQs), we evaluate the effectiveness of QECK for improving the performance of code search algorithms, and investigate the performance of *QECK_Rocchio*. We conduct three experiments to answer the three RQs, respectively. We also provide the details about the data set employed in our experiments and the evaluation for the results. Our experiments are conducted on a 3.60 GHz CPU (Intel i5) PC running windows 8.1 OS with 8G memory. We implement *QECK_Rocchio* using Java 1.7.0 in Eclipse.

### 4.1 Research Questions

We explore the following three RQs:

*RQ1: Whether QECK can improve the performance of code search algorithms?*

In our first experiment, we utilize three code search algorithms [21], [27] to verify the effectiveness of QECK by comparing the performance of recommendation before and after QECK is applied. From comparative results of these code search algorithms, we want to identify whether or not our QECK technique can indeed improve the retrieval performance.

Specifically, the first code search algorithm is BM25 based information retrieval approach on Lucene, which is denoted as IR in the first experiment. The second is Portfolio [27] based on Vector Space Model (VSM), PageRank, and Spreading Activation Network (SAN). The third is VF [21] based on VSM and the frequent item-set mining. We program the IR, and reproduce Portfolio and VF by following the parameters setting and the steps claimed in literatures [21], [27].

In order to conduct fair comparison, we optimize the parameters, the number of PRF documents and the number of expansion words, as follows. Based on the recommendation found in the domain literatures [10], [17], we first set the top five Q&A pairs returned in the first-pass retrieval as the PRF documents. Then, we adjust the number of expansion words over (n ∈ (1 − 10,15,20)) to achieve the best performance for each code search algorithm, respectively.

*RQ2: How the parameters affect the performance of QECK?*

To further study the influence of parameters variation on performance of QECK, we conduct the second experiment. Two factors influencing QECK are the quality of Q&A pairs and the expansion words selected from these Q&A pairs, which correspond to two parameters: the number of PRF documents (i.e. feedback Q&A pairs) and the number of expansion words, respectively.

As it is a time-consuming task to label relevant scores for code snippets, we only discuss the situation when fixing a parameter and adjusting another parameter. Specifically, first, we set the top five Q&A pairs as the PRF documents, and explore the trend of performance for each code search algorithm by varying the number of expansion words. Based on the trend, we achieve the optimal value of the number of expansion words for each code search algorithm. Then, we fix this optimal value, and explore the trend of performance for each code search algorithm by varying the number of PRF documents.

TABLE 1
QUERIES FOR TEST

| ID | Query | Tags | Viewed times |
|----|-------|------|--------------|
| 1 | Record audio sound | android | 4783 |
| 2 | Get screen dimensions in pixels | android, layout, screen | 720031 |
| 3 | Take a screenshot in Android | android, screenshot | 107071 |
| 4 | Get the memory used | android, memory, memory-management | 217026 |
| 5 | Get the list of activities/applications installed | android | 180820 |
| 6 | Import the system time | android, operating-system | 36113 |
| 7 | Open a URL in Android's web browser | android, url, android-intent, android-browser | 342424 |
| 8 | Use android Timer in Android activity | android, multithreading, timer, scheduled-tasks | 18998 |
| 9 | Capture Image from Camera and Display in Activity | android, image, camera, capture | 157947 |
| 10 | Handle right to left swipe gestures | android, swipe, gesture-recognition | 152674 |
| 11 | Converting pixels to dp | android | 264672 |
| 12 | Draw a line in android | android | 182837 |
| 13 | Get cpu usage | android, cpu-usage | 72209 |
| 14 | Detect network connection status | android, networking, wifi, connectivity | 69553 |
| 15 | Check if an application is installed or not in Android | android, apk | 46174 |
| 16 | Convert an image into Base64 string | android, base64 | 80049 |
| 17 | Get the web page contents from a WebView | android, android-webview | 52124 |
| 18 | Cancel an executing AsyncTask | android, android-asynctask | 85973 |
| 19 | Detect if a Bluetooth device is connected | android | 39245 |
| 20 | Retrieve incoming call's phone number | android, telephonymanager, phone-state-listener | 50134 |

*RQ3: Whether our code search approach, QECK based Rocchio's model, is better than the state-of-the-art method?*

As the state-of-the-art method, Lu et al. [28] present a solution for improving the code search based on query reformulation technique, denoted as $P_{WordNet}$ in the experiment. To implement $P_{WordNet}$, the authors, first, find the synonyms of each word in the given query with the same POS using WordNet to expand the queries. Then, they identify the key source code identifies in methods as the keyterms based on the similarity values with the expanded queries. Finally, according to the percentage of keyterms in each method, they recommend top-k methods to developers.

We compare our approach $QECK_{Rocchio}$ against $P_{WordNet}$ in the same retrieval scenario, and verify the effectiveness of our approach. The parameters in each of these two approaches are adjusted to achieve the best performance, respectively.

### 4.2 Dataset Collection

In this part, we describe the query set, the Q&A collection, and the code snippet corpus used in three experiments. These datasets can be found in our webpage [5].

#### 4.2.1 Query Set

In our experiments, we employ 20 programming tasks to form the original queries. These tasks are real-world programming tasks collected from Stack Overflow [7].

For collecting these programming tasks, we first manually rank the posts with the "android" tag on Stack Overflow. Then, we check the posts one by one based on some criteria until we collect 20 tasks. The criteria [27] are that the tasks should belong to the framework of Android app development and be viewed several times. Meanwhile, there are solutions along with these programming tasks in the webpages, as these solutions can provide assists for evaluating the relevance scores of code snippets.

Each programming task contains a title and a description. We find that the structure of descriptions are different for different tasks. Some contain only textual description, whereas some contain both code context and text description. In our experiments, for each of 20 programming tasks, we simply extract the words in the title as the query. For example, for the programming task in the Fig. 2 (a), we extract the words "*take a screenshot in Android*" as the query. In this process, the extracted words may not always represent actual developer queries as some words in the description may be neglected. This will be a threat to our results.

Table 1 shows 20 queries. The column "Tags" shows the categories of these queries. Note that our queries belong to different categories. The column "Viewed Times" indicates the number of times a query has been viewed by visitors. These values are all comparatively large, which means that the developers desire to achieve the solutions of these programming tasks.

#### 4.2.2 Q&A Pair Collection

We download a release of Stack Overflow public data dump (the version of August 2015), *posts.xml*. This dataset stores all questions and their answers (denoted as posts) on Stack Overflow until the dump is built. Following the steps in Section 3.2, we construct 5,108,770 Q&A pairs from total of 24,120,522 posts in this dump. Finally, we achieve a collection containing 312,941 Q&A pairs that are labeled with the "android" tag. Notably, to avoid introducing bias, in Q&A pair collection, we remove the Q&A pairs corresponding to our programming tasks.

#### 4.2.3 Code Snippet Corpus

Following the steps in Section 3.4, until August 2015, we collect totally 1,538 Android app projects from F-droid. By segmenting the Java files in these projects, we construct a corpus containing 921,713 code snippets.

## 4.3 Evaluation

### 4.3.1 Steps of Evaluation

After achieving the recommendation results from three experiments, we need to manually label each code snippet in the results, and calculate the metrics for each comparative algorithm.

Specifically, we set up the evaluation as follows [62]. For a given query, first, we obtain the Top-k (k=10 in our study) code snippets returned by each comparative algorithm in three experiments. Then, we merge all code snippets into a pool, which includes only unique code snippets. For each code snippet in this pool, we recruit two participants to label the relevance scores with Four-level Likert scale. Participants judge the relevance scores by the solutions appearing with the programming tasks on the webpages and their programming experience. As regards the inconsistencies of labeling, we recruit an expert to arbitrate the score. Finally, based on the scores, we exploit two popular metrics to inspect the performance of each algorithm. In the whole process, about which algorithm is a code snippet from, it is invisible for the participants and expert as the code snippets are merged into a pool. Meanwhile, two participants and the expert could view the queries and the corresponding programming tasks on Stack Overflow.

Two participants are graduate students from Dalian University of Technology who have at least four years of Java experience. Moreover, the expert, one author of this paper, is a doctoral student who has more than nine years of Java experience. Both two participants and the expert have at least three years of Android application development experience. Before the labeling process, we give them a 30-minutes training about labeling.

The guidelines of labeling code snippets are as follows [27]:

Score 4: Highly relevant. The code snippet is perfectly suitable for the programming task.

Score 3: Mostly relevant. The code snippet or the APIs contained in this snippet can be reused for the programming task with some changes.

Score 2: Mostly irrelevant. The code snippet only contains a little relevant code lines, which is not enough to solve the programming task.

Score 1: Completely irrelevant. The code snippet cannot solve the programming task.

In simple terms, a code snippet labeled with 3 or 4 means that this code snippet should contain the useful code lines or APIs to solve the programming task.

Totally, two participants label 3251 code snippets in our three experiments. Among these code snippets, two participants label same scores for 2756 code snippets. The degree of consensus is 84.8%.

### 4.3.2 Metrics

An ideal recommendation algorithm should hit more of the relevant answers and place them at the top of the results. Following the previous work [27], we adopt two types of metrics to evaluate the performance of each algorithm in three experiments, namely, Precision and Normalized Discounted Cumulative Gain (NDCG). In this paper, the performance of algorithms refers to the effectiveness of algorithms. We also discuss the efficiency of algorithms in Section 6. For comparing the performance of different algorithms, following [21], [27], we calculate the mean value of two metrics for the queries.

Specifically, the Precision@K is defined as the proportion of the true positives (i.e. the code snippets with score 3 or 4) in Top-k recommended results (both true positives and false positives) for a given query [27].

The Precision@K is calculated as:

$$Precision@K = \frac{(|Relevance|)}{(|Retrieved|)} \qquad (6)$$

Where the denominator $|Retrieved|$ is the total number of results recommended by an algorithm, which equals to 10 in our study. The numerator $|Relevance|$ is the number of relevant code snippets in the result.

NDCG is commonly used in the information retrieval to measure the ranking capability of a recommendation algorithm. A algorithm is more useful when there are more relevant results in higher positions in the hit list than irrelevant results. We calculate NDCG@K of each algorithm for a given query, as:

$$NDCG@K = \frac{DCG@K}{IDCG@K} \qquad (7)$$

$$DCG@K = R_1 + \sum_{i=2}^{K} \frac{R_i}{log_2 i} \qquad (8)$$

where $NDCG@K$ is the $DCG@K$ normalized by $IDCG@K$. $IDCG@K$ is the ideal $DCG@K$, where the results are sorted by relevance scores. $R_1$ is the relevance score at the first position in the list. $R_i$ is the relevance score at the $i$-$th$ position.

Notably, in the experiments, we observe that the value of NDCG cannot show the real performance for recommendation. For example, there are two results from two algorithms for a given query, respectively. The result A is 4, 1, 1, and 1. The result B is 2, 2, 2, and 2. Here, we recommend Top-4 code snippets in the results. The values of NDCG for two results are all equal to 1. However, we find that there are no relevant code snippets for the given query in the result B, as a snippet with score 3 or 4 is considered to be relevant. To solve this problem, we set the score 1 and score 2 to score 0 in the returned results. Then, the NDCG value of the result B equals to 0. The NDCG value of the result A still equals to 1.

TABLE 2
THE STATISTICAL SUMMARY OF THREE CODE SEARCH ALGORITHMS (IR, PORTFOLIO, AND VF) BEFORE AND AFTER QECK IS APPLIED

| Metrics | Approach | Samples | Min | Max | Median | Mean | StdDev |
|---------|----------|---------|-----|-----|--------|------|--------|
| Precision | IR | 20 | 0% | 100% | 60% | 57.5% | 0.2552 |
| | $IR_{QECK}$ | 20 | 30% | 100% | 85% | **79.5%**+(0.007) | 0.2164 |
| | Portfolio | 20 | 0% | 90% | 60% | 55.5% | 0.2350 |
| | $Portfolio_{QECK}$ | 20 | 10% | 100% | 80% | **74%**+(0.014) | 0.2664 |
| | VF | 20 | 0% | 100% | 50% | 47% | 0.3278 |
| | $VF_{QECK}$ | 20 | 20% | 100% | 90% | **77%**+(0.013) | 0.2774 |
| NDCG | IR | 20 | 0 | 1 | 0.7772 | 0.7551 | 0.2407 |
| | $IR_{QECK}$ | 20 | 0.4864 | 1 | 0.9461 | **0.9030**+(0.033) | 0.1271 |
| | Portfolio | 20 | 0 | 1 | 0.7795 | 0.7445 | 0.2325 |
| | $Portfolio_{QECK}$ | 20 | 0.3562 | 1 | 0.9328 | **0.8661**+(0.020) | 0.1761 |
| | VF | 20 | 0 | 0.9816 | 0.8220 | 0.6347 | 0.3526 |
| | $VF_{QECK}$ | 20 | 0.5089 | 1 | 0.9221 | **0.8599**+(0.033) | 0.1493 |

*The "+" refers to the p-value is less than 0.05 among the pairwise comparison for each algorithm. The p-values are surrounded by parentheses. The better mean values among the pairwise comparisons are shown in bold font.*

## 5 RESULTS AND ANALYSIS

The section provides three experimental results to answer the RQs mentioned in Section 4.1, respectively.

In the first and third experiment, for drawing confident conclusions whether one recommendation result outperforms another, we conduct a statistical test to compare the mean values of two metrics (i.e. Precision and NDCG) for two results. Specifically, we conduct the two-sided Wilcoxon's signed rank test between two results. When comparing each pair of results, the primary null hypothesis is that there is no statistical difference in the performance between two results. In this section, we adopt the 95% confidence level (i.e. the p−values below 0.05 are considered significant).

### 5.1 RQ1

The first experiment is conducted to identify the effectiveness of QECK for improving the retrieval performance.

Table 2 contains the details of extremal values, median, mean, and standard deviation of Precision and NDCG, also shows the performance comparisons of three code search algorithms before and after QECK is applied. In this table, the "+" refers to that the p-value is less than 0.05 among the pairwise comparison for each algorithm. From this table, we can observe that the p-values are all less than 0.05 among the pairwise comparisons. For each code search algorithm before and after QECK is applied, we reject the null hypothesis and accept the alternative hypothesis that there is a statistically significant difference in the mean values of Precision and NDCG, respectively.

Fig. 3 shows the statistical summary of the performance comparisons of two metrics for three code search algorithms. In Table 2 and Fig. 3, we can observe that, after using QECK, the performance is improved for each algorithm. Specifically, for the mean value of Precision, the improvement of IR is 38%, Portfolio is 33%, and VF is 64%. For the mean value of NDCG, the improvement of IR is 20%, Portfolio is 16%, and VF is 35%. The foremost reason for these improvements is the usage of the software-specific words that are extracted from PRF Q&A pairs. QECK increases the possibility of searching more relevant code snippets.

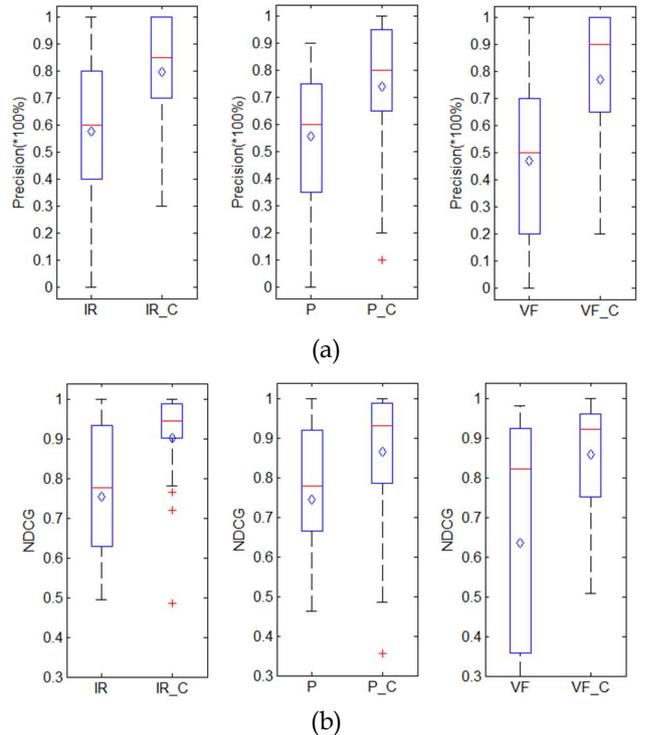

Fig. 3. The statistical results of Precision (a) and NDCG (b) for three code search algorithms (IR, Portfolio (P), VF) before and after QECK is applied. Here, IR_C refers to QECK based IR, P_C refers to QECK based Portfolio, VF_C refers to QECK based VF. The x axes indicate two code search algorithms, respectively. The y axes indicate the range for two metrics, respectively. The red line represents the median. And the blue rhombus represents the mean.

***Answer RQ1***: After using our proposed QECK technique, query expansion based on crowd knowledge, the performance of three code search algorithms are all improved.

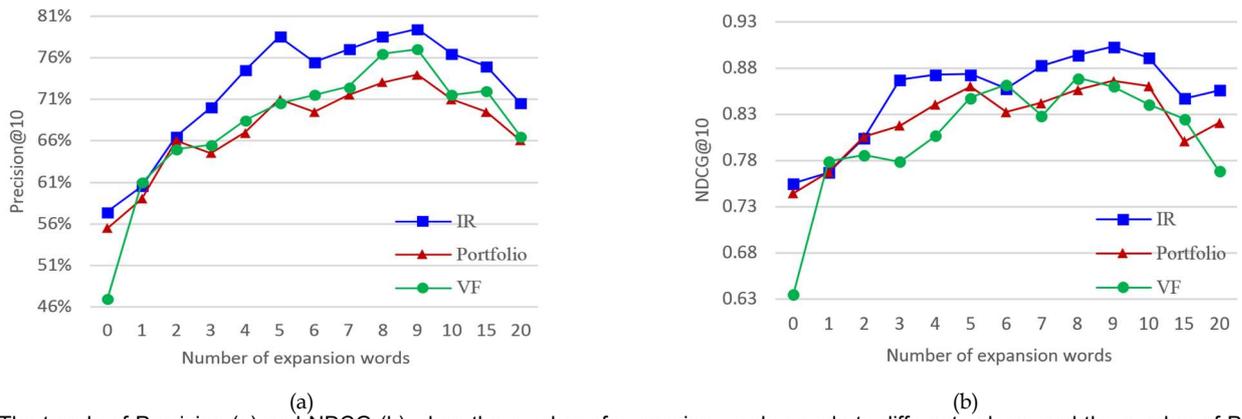

(a)                                                                                        (b)

Fig. 4. The trends of Precision (a) and NDCG (b) when the number of expansion words equals to different values, and the number of PRF documents is set as 5.

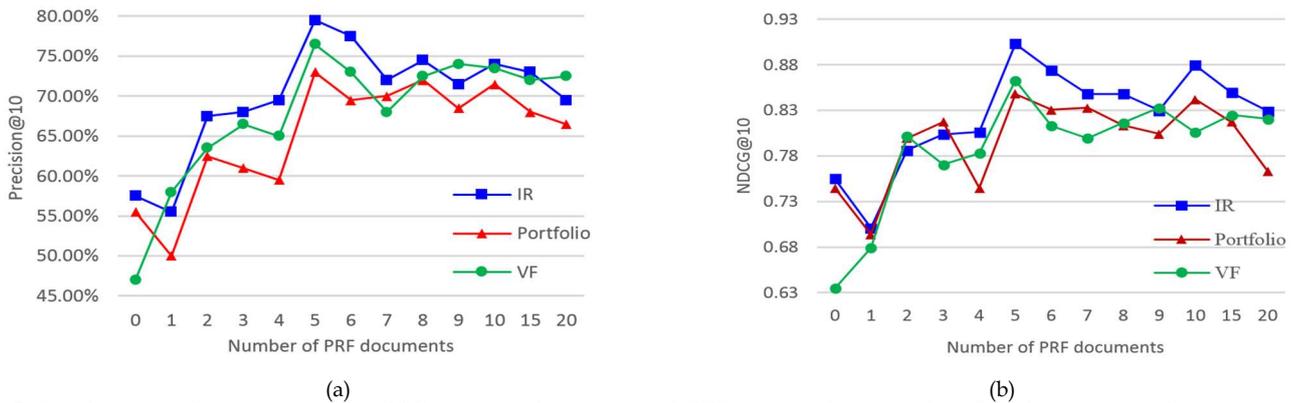

(a)                                                                                        (b)

Fig. 5. The trends of Precision (a) and NDCG (b) when the number of PRF documents equals to different values, and the number of expansion words is set as 9.

## 5.2 RQ2

We conduct the second experiment to further explore the influence of parameters variation on performance of QECK. Following the first experiment, we compare the performance trends of three code search algorithms by changing the number of PRF documents and the number of expansion words.

Fig. 4 and Fig. 5 show the trends of three code search algorithms before and after QECK is applied by means of two metrics (Precision@10 and NDCG@10) when we fix a parameter and change another. In the subfigures of two figures, three curves refer to three code search algorithms, respectively. The value "0" on the horizontal axes refers to the situation using original query. The other values on the horizontal axes refer to the situations using query expansion when selecting different values of the number of PRF documents or the number of the expansion words. Fig. 4 shows that the performance of three code search algorithms reach the optimum state when the number of expansion words equals to 9, and the number of PRF documents is fixed as 5. In contrast, Fig. 5 shows that the performance of three code search algorithms almost reach the optimum state when the number of PRF documents equals to 5, and the number of expansion words is fixed as 9.

Here, we show and analyze some interesting findings in two figures:

- After employing our QECK technique, the performance of three algorithms are generally better. This indicts that our proposed technique can make use of Q&A pairs and add useful expansion words to generate more useful expansion queries.

- The performance of all code search algorithms increase at the beginning when the value of each of two parameter grows up. There is a unique optimal value for each of two parameters. This trends can be explained by considering the percentage of truly useful documents and expansion words in the PRF documents and the adding expansion words, respectively [9]. Taking the number of expansion words as example, on the one hand, if we select a very small number of expansion words, it is more likely that we will get little useful words for some queries. On the other hand, if we select a larger number of expansion words, it is more likely that some irrelevant words will be added, and the performance deteriorates [42].

- The performance of IR is better than that of Portfolio and VF in most cases for two metrics. Therefore, we consider QECK based IR method (i.e. QECK based Rocchio's model) as our total solution to search code.

Based on the findings and analysis above, we recommend that, in QECK, the default value for the number of PRF documents is 5, and the default value

TABLE 3
THE STATISTICAL SUMMARY OF TWO METHODS

|  | Approach | Samples | Min | Max | Median | Mean | StdDev |
|---|---|---|---|---|---|---|---|
| Precision | $P_{WordNet}$ | 20 | 30% | 90% | 65% | 65% | 0.2037 |
|  | $QECK_{Rocchio}$ | 20 | 30% | 100% | 85% | 79.5% | 0.2109 |
| NDCG | $P_{WordNet}$ | 20 | 0.3992 | 0.9818 | 0.7688 | 0.7794 | 0.1543 |
|  | $QECK_{Rocchio}$ | 20 | 0.4864 | 1 | 0.9461 | 0.9030 | 0.1239 |

for the number of expansion words is 9.

*Answer RQ2:* When fixing the number of PRF documents and varying the number of expansion words, there is a unique optimal value of performance for each code search algorithm. Too many or too less expansion words is not desirable.

### 5.3 RQ3

The third experiment is conducted to compare our code search method $QECK_{Rocchio}$ against $P_{WordNet}$ [28]. Table 3 shows the details of extremal values, median, mean, and standard deviation of Precision and NDCG, and presents the performance comparisons of two methods.

We can observe that the p-values are all less than 0.05 among the pairwise comparisons for two metrics. Then, we reject the null hypothesis and accept the alternative hypothesis that there is a statistically significant difference in the mean values of Precision and NDCG for $QECK_{Rocchio}$ and $P_{WordNet}$, respectively.

Fig. 6 shows the statistical summary of the performance comparisons of two metrics for two methods. In Table 3 and Fig. 6, we can observe that the performance of $QECK_{Rocchio}$ is better than $P_{WordNet}$. Specifically, for Precision, the improvement is 22% (p-value = 0.045). For NDCG, the improvement is 16% (p-value = 0.005). The results further verify our conclusion that the utilization of software-specific words from QECK is effective for code search based on query expansion.

*Answer RQ3:* Based on the observations and analysis above, we can argue that our method $QECK_{Rocchio}$ is a better method for code search than the state-of-the-art method. This result clearly validates the ability of QECK for providing useful expansion words.

## 6 THREATS TO VALIDITY

This section discusses the threats to the validity of our work. We list them as follows.

*Labeling:* As mentioned in Section 4.3, in the process of labeling the relevance scores for code snippets, we recruit two participants. They have three years of experience in Android app development, and four years of experience in Java development. We believe that they can assign the correct relevance scores with the help of their experience and the solutions that appearing with the programming tasks in the webpage. However, they still have different programming levels and may label the same code snippet with different scores. This threat is minimized by recruiting an expert to arbitrate the score.

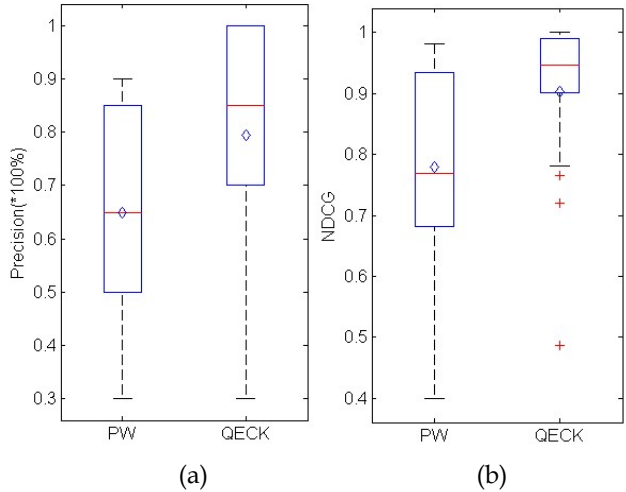

Fig. 6. The statistical results of Precision (a) and NDCG (b) for two query expansion based code search methods: our method $QECK_{Rocchio}$ denoted as QECK, and WordNet based method ($P_{WordNet}$) denoted as PW. The x axes indicate two code search algorithms, respectively. The y axes indicate the range for two metrics, respectively. The red line represents the median. And the blue rhombus represents the mean.

*The query set:* In our three experiments, we use 20 queries to evaluate our technique. The size is the same as the previous researches [8], [21]. However, it may still threaten the validity of the conclusion. We will test more queries in our further work. Moreover, in the process of generating the query set, we simply extract the words in the titles of programming tasks as the queries. Obviously, some words in the descriptions will be neglected, which means that queries may not be representative of actual developer queries. This will be a threat to validity. In our future work, we will employ the whole programming task (i.e., the title and description) as the query, and explore the impact of the description on results.

*The Q&A collection and the Code snippet corpus:* In three experiments, we return PRF documents from a Q&A collection in the first-pass retrieval of $QECK_{Rocchio}$, and retrieve code snippets on a code snippet corpus in the second-pass retrieval. Two datasets are all real-world, and have a certain scale. Based on two datasets, the effectiveness of our QECK technique has been demonstrated. However, these datasets are just about Android, and the scale is still smaller than other sets [1], [3], [6]. In future work, we plan to explore the effectiveness of QECK on larger datasets.

*The Comparative methods:* We reproduce two code search algorithms [21], [27] in our first and second experiments and a comparative method [28] in our third experiment. There are certain gaps in the performance between our reproduced methods and the original ones. The reasons may be the differences of queries or the code snippet corpora. Specifically, our queries are all about Android app development, and the corpus is collected from Android app projects rather than traditional software projects.

TABLE 4
COMPARISON BETWEEN QECK$_{Rocchio}$ WITH RELATED METHODS

| | | Code Search: Input Type | | | | |
|---|---|---|---|---|---|---|
| | | Test Case [22],[23] | Keywords or Free-form queries [8], [21], [27] | Others [19], [48], [61] | | QECK$_{Rocchio}$ |
| Query Expansion [10], [26], [58] | Global | Lemos et al.[24]: WordNet, code-related thesaurus | Lu et al. [28]: POS, WordNet | | | Local; PRF; Q&A Pairs on SO; |
| | Local | | Our Method | | | |
| | | | Wang et al. [56]: users' feedback | | | |
| | | | Hill et al. [38]: most highly co-occurring words | | | |

*Parameters:* In our experiments, as it is a time-consuming task for labeling code snippets, we only discuss the situation when fixing a parameter and adjusting another parameter. Furthermore, we arbitrarily add the expansion words into original queries to generate the expansion queries. It means that the weights of the words in the original queries and the expansion words are viewed as equal. Actually, different weights may produce different recommendation performances [10]. In our future work, we plan to automatically adjust two parameters and the weights of words to achieve better performance.

*Efficiency:* In this paper, we focus on improving the effectiveness of recommendation method. Similar to the previous studies [21], [27], we employ two metrics (i.e., Precision and NDCG) to measure the effectiveness. However, the efficiency of a recommendation method is also crucial, which depends on the quality of code, the size of data, and the algorithm itself. Currently, our method finishes a recommendation within 3 seconds averagely. This will be a threat for using in practice. In the future work, we will improve the quality of code and build a plug-in for Eclipse.

## 7 RELATED WORK

This section discusses the literatures related with our research, which involves two aspects: the researches about code search based on query expansion and about Q&A pairs recommendation on Stack Overflow.

### 7.1 Query Expansion based Code Search

In this part, we first introduce the studies about code search and query expansion, and explain the relationship between our method and two types of researches as shown in Table 4. Then, we show the researches about code search based on query expansion, and the differences between these researches and our work.

For the studies about code search, according to input types, there are several categories: test cases, free-form queries, and others. Specifically, Lemos et al. [22], [23] propose CodeGenie to perform the code search by employing queries generated from the information (i.e., names of classes and methods, and interfaces) available on test cases. Except the test cases, free-form queries represented with several words are employed as the inputs to retrieve a ranked list of relevant code snippets [8], [21], [27]. Moreover, other kinds of input for code search contain: the names of API methods [30], [33], [61], the pairs of class types [25], the examples of desired code [48], the pairs with source and destination types [50], the structural context [19], [40], the code under editing [14], [32], and so on. Our work focusses on the studies taking free-form queries as input.

For the studies about query expansion, there are two major classes: global approaches and local approaches [10], [26], [58]. Specifically, the global approaches reformulate queries with a thesaurus, like WordNet, by leveraging related words and synonyms from this thesaurus. In contrast, based on the documents initially appearing to match the original queries, the local approaches leverage the user' marks on these documents (Relevance Feedback, RF) or automatically extract expansion words (Pseudo Relevance Feedback, PRF) from these documents. Our method employs the pseudo relevance feedback, one of the local approaches, to expand queries.

As query expansion can solve the term mismatch problem, it has been shown to be effective on many natural language processing (NLP) tasks [10], [26], [58]. For improving the performance of code search, some query expansion based methods are presented in recent years. For example, Wang et al. [56] integrate the users' feedbacks to make the more relevant code snippets appearing earlier in the list. However, this work need human intervention. For automating this process, Hill et al. [18], [38], [44] propose a source code search technique, CONQUER, which refines the queries by suggesting the most highly co-occurring words that appearing in the source code as alternative query words. In 2015, Lu et al. [28] provide a query reformulation technique, which is denoted as $P_{WordNet}$ in our experiment, based on part-of-speech of each word in queries and WordNet. The results of $P_{WordNet}$ show that it can help recommend good alternative queries, and outperform CONQUER. In our study, we employ $P_{WordNet}$ as a comparative method in our third experiment. Lemos et al. [24] present an automatic query expansion approach, AQE, which uses test cases as inputs, and leverages WordNet and a code-related thesaurus [59] to expand queries.

These approaches mentioned above either use test cases as inputs rather than free-form queries, or have not employed the software-specific words to expand queries. Different from these researches, in our study, QECK automatically extracts software-specific expansion words from PRF Q&A pairs on Stack Overflow, and takes the free-form queries as the inputs.

## 7.2 Q&A Pairs Recommendation

Many studies have been conducted on solving the software engineering tasks by leveraging the discussions on Stack Overflow, for example, mining source code description [54], extracting cookbooks for APIs [12], helping developers debugging their code [11], assisting software comprehension and development [36], [46], [37], locating method definitions on Stack Overflow [55], and so on. We focus on the studies about recommending Q&A pairs on Stack Overflow.

In 2013, Ponzaneli et al. [36] present Seahawk, which automatically generates queries by extracting words from the code entities in the integrated development environment (IDE), and displays the Q&A pairs on Stack Overflow to developers for software comprehension. Seahawk employs the textual similarity between queries and Q&A documents. Based on Seahawk, Ponzanelli et al. [37] propose a tool, Prompter, which automatically retrieves discussions on Stack Overflow by using several combined aspects. Souza et al. [46] improve Seahawk by integrating the textual similarity and the scores of Q&A pairs that are voted by crowd. They combine two types of scores by performing a normalization step to generate the final score. Using the same strategy like [46], we retrieve PRF Q&A pairs for given queries in our paper. The different is, to assure the quality of Q&A pairs, we only select the answers with the label "AcceptedAnswer" rather than each answer.

## 8 CONCLUSION AND FUTURE WORK

For improving the performance of code search, in this paper, we propose query expansion based on crowd knowledge (QECK) to solve the vocabulary problem. To evaluate the effectiveness of QECK, we explore three Research Questions (RQs) in three experiments on real-world datasets on the Android platform. The results of three experiments state that: first, after using QECK, the performance of three code search algorithms are improved by up to 64% in Precision, and 35% in NDCG. Second, too many or too less expansion words are not desirable for code search based on query expansion. Third, comparing to the state-of-the-art query expansion method, our method QECK$_{Rocchio}$, the implementation of QECK in the Rocchio's model, improves 22% for Precision and 16% for NDCG. These results verify the effectiveness of QECK for code search in aiding mobile app development. It also means that the utilization of software-specific words in QECK is effective for code search.

We consider two aspects as our future work. First is analyzing queries based on its features, which includes automatically assessing performance of a query [16], [35] and automatically recommending a reformulation strategy for a given query [17]. Second, for improving the performance of the automatic query expansion, there are some efforts, for example, heuristic term frequency transformation model [60] to capture the local saliency of a candidate term in the feedback documents, general solution to improve the efficiency of pseudo relevance feedback methods [57], and so on. These techniques could be employed by our QECK technique to further enhance the effectiveness.

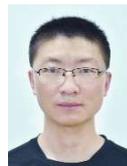
Liming Nie received the M.Sc. degree in Guangxi University for Nationalities, Nanning, China, in 2009. He is currently a Ph.D. candidate in Dalian University of Technology. His current research interests include code recommendation and mining software repositories in software engineering.

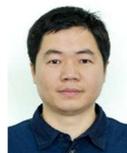
He Jiang received the Ph.D. degree in computer science from the University of Science and Technology of China, Hefei, China. He is currently a Professor with the Dalian University of Technology, Dalian, China. His current research interests include search based software engineering and mining software repositories. Dr. Jiang is also a member of the ACM and the CCF.

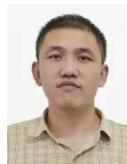
Zhilei Ren received the B.Sc. degree in software engineering and the Ph.D. degree in computational mathematics from the Dalian University of Technology, Dalian, China, in 2007 and 2013, respectively. He is currently a lecturer with the Dalian University of Technology. His current research interests include evolutionary computation and its applications in software engineering. Dr. Ren is a member of the ACM and the CCF.

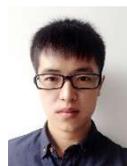
Zeyi Sun received the B.Sc. degree in software engineering from the Dalian University of Technology, Dalian, China, in 2015. He is currently a Master of Software Engineering candidate in Dalian University of Technology. His current research interest is code recommendation in software engineering.

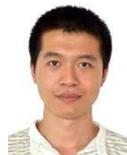
Xiaochen Li received the M.Sc. degree in software engineering from the Dalian University of Technology, Dalian, China, in 2015. He is currently a Ph.D. candidate in Dalian University of Technology. His current research interest is mining software repositories in software engineering.